\documentclass[twocolumn,showpacs,prl]{revtex4} 
\usepackage{graphicx}
\begin{document}  
\title{$^{29}$Na: Defining the edge of the Island of Inversion for $Z=11$ } 
\author{Vandana Tripathi$^1$, S.L. Tabor$^1$, P.F. Mantica$^{2,3}$, C.R. 
Hoffman$^1$,  M. Wiedeking$^1$, A.D. Davies$^{2,4}$,  S.N. Liddick$^{2,3}$, 
W.F. Mueller$^2$, T. Otsuka$^{5,6}$, A. Stolz$^2$, B.E. Tomlin$^{2,3}$, Y. 
Utsuno$^7$, A. Volya$^1$}
\affiliation{\mbox {$^1$Department of Physics, Florida State University, 
Tallahassee, Florida 32306, USA} 
\mbox {$^2$National Superconducting Cyclotron Laboratory, Michigan State 
University, East Lansing, Michigan 48824, USA}
\mbox {$^3$Department of Chemistry, Michigan State University, East Lansing, 
Michigan 48824, USA}
\mbox {$^4$Department of Physics and Astronomy, Michigan State University, 
East Lansing, Michigan 48824, USA}
\mbox{$^5$Dept. of Physics and Center for Nuclear Study, University of Tokyo, 
Hongo, Tokyo 113-0033, Japan}
\mbox{$^6$ RIKEN, Hirosawa, Wako-shi, Saitama 351-0198, Japan }
\mbox{$^7$Japan Atomic Energy Research Institute, Tokai, Ibaraki 319-1195, 
Japan }
}
\date{\today}
\begin{abstract}
The low-energy level structure of the exotic Na isotopes $^{28,29}$Na has been 
investigated through $\beta$-delayed $\gamma$ spectroscopy. The $N=20$ isotones
 for $Z=10-12$ are considered to belong to the ``island of inversion'' where 
 intruder configurations  dominate the ground state wave function. However, 
it is an open question as to where and how the transition from normal to 
intruder-dominated configurations happens in an isotopic chain. The present 
work, which presents the first detailed spectroscopy of $^{28,29}$Na, clearly 
demonstrates that such a transition in the Na isotopes occurs between $^{28}$Na
 ($N=17$) and $^{29}$Na ($N=18$), supporting the smaller $N=20$ shell gap in 
neutron rich {\it sd} shell nuclei. The evidence for inverted shell structure
is found in $\beta$ decay branching ratios, intruder dominated spectroscopy of
low-lying states and shell model analysis. 
\end{abstract}
\maketitle

Shell structure is one of the most fundamental properties of atomic nuclei that
emphasizes the mean-field nature of nucleon dynamics. The spherical shells 
characterize the most stable configurations, the magic nuclei.  However, from 
the pioneering studies of Thibault {\it et al.}, \cite {thibault}, it became 
apparent that the magic numbers may not be global and the $N=20$ magic number 
may disappear in neutron-rich nuclei around $Z=11$. The anomalously large 
binding energies for Na isotopes near $N=20$ \cite {thibault}, the observation 
of low-lying first excited states with large B(E2) transition probabilities to 
their ground states in $^{32}$Mg \cite {moto, prity, chiste}, $^{28,30}$Ne 
\cite {prity, Yana}, and $^{31}$Na \cite {prity2} and the measurements of 
large quadrupole moments in neutron-rich Na isotopes \cite {keim} are clear 
experimental signatures of this effect. The cause for this unexpected behavior,
 or ``inversion'' as it is called \cite {warburton, caurier}, is due to the 
competition between the {\it normal} and {\it intruder} configurations 
(excitations across the $N=20$ shell gap) with the intruder configurations 
becoming energetically favored. This emphasizes the phase transition from 
nucleon motion in the mean field to dynamics dominated by the residual 
nucleon-nucleon interactions.

For stable nuclei, the $N=20$ shell gap separating the {\it sd} and {\it fp} 
shells is about 6 MeV, making it energetically costly for nucleons to occupy 
the {\it fp} orbitals. The ``inversion'' observed in neutron rich nuclei is 
related to the varying gap between the $d_{3/2}$ and $f_{7/2}$ orbitals which 
occurs due to the strong attractive interaction between the neutrons and 
protons occupying the spin-orbit coupling partners ($j_>$ and $j_<$) \cite 
{otsuka}. In exotic nuclei, due to the neutron-proton unbalance, the $j_<$ 
orbit is located rather high, reducing the gap with the higher shell. This 
reduced shell gap ($\sim$3 MeV) in the presence of residual nucleon-nucleon 
interactions is not enough to sustain a shell structure. The transition from 
the normal to an intruder ground state depends primarily on the competition  
between shell structure and many body correlation effects due to residual 
interactions, which is a sensitive function of neutron number \cite 
{otsukanew}. The open shell nuclei with $N<20$ are very sensitive to the shell 
gap, as the gain in correlation energy for the intruder is not as large as for 
closed shell nuclei. The presence of intruder configurations at low excitation 
energy in transitional nuclei would be a signature of a small $N=20$ shell gap.
 Recently, large scale Monte Carlo Shell Model (MCSM) calculations by Utsuno 
{\it et al.,} \cite {utsuno, utsuno2} showed the intricate relation between the
melting of shell structure and stability of exotic nuclei. The transition to an
 inverted state relates to the general question of phase changes in small 
systems, which unlike sharp changes in the thermodynamic limit exhibit gradual 
transitions accompanied by regions dominated by large fluctuations; furthermore
 particle number becomes of central importance. 

In the present work we address this issue by approaching the $N=20$ boundary 
along the Na isotopes.  We have performed detailed $\beta^-$ delayed 
$\gamma$-spectroscopy measurements of $^{28,29}$Na  ($N=17,18$) to investigate 
the transition from normal-dominant to intruder-dominant states in the chain of
 Na isotopes and its connection with the predicted small $N=20$ shell gap for 
these neutron rich nuclei. Prior to the present work, very limited experimental
 information on excited states in these nuclei was available \cite {reed}. The 
level structures obtained from the present study clearly demonstrate the 
dominance of excitations across the $N=20$ shell at low energies for $^{29}$Na 
but a nearly pure {\it sd} shell configuration for $^{28}$Na. This defines the 
extreme edge for the regime of intruder domination in Na isotopes, below which 
the intruder configurations cannot compete with normal configuration at small 
excitation.  

Earlier comparisons of the experimental masses of the Na isotopes 
to shell model results within the {\it sd} shell (USD)\cite {alex} 
suggest that the ``inversion'' occurs sharply at $N=20$. However comparison of 
the electric and magnetic moments of the Na isotopes \cite {keim} indicates a 
more gradual change; for the $N=19,20$  isotopes, the moments cannot be 
reproduced by the USD model, while the measured quadrupole moment of $^{29}$Na 
($N=18$) is 30\% larger than predicted by the USD model, and could be explained
 by considering $\sim$42\% mixing of intruder configurations in the ground 
state of $^{29}$Na \cite {otsukanew}. The spectrum of excited states provides 
another sensitive tool to probe the shell gap as well as the mixing between 
normal and intruder configurations. In particular, the relatively small 
$d_{3/2}$ - $f_{7/2}$ shell gap estimated at about 3.3 MeV in recent 
calculations \cite {otsukanew} for $^{29}$Na can be tested by examining the 
low-energy excited states. 

\begin{figure}
\begin{center}
\includegraphics[scale=0.75]{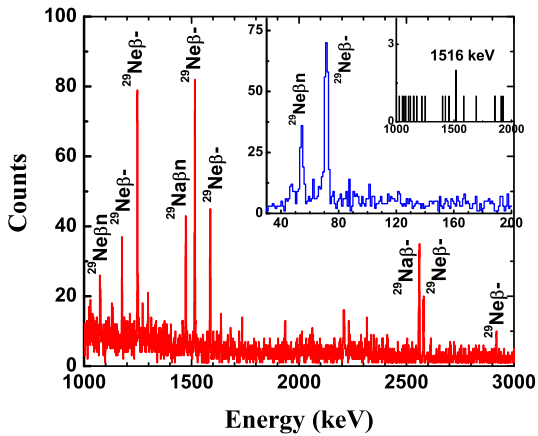}
\caption{(Color online) $\beta^-$ delayed $\gamma$-ray spectrum in the range 
1.0 - 3.0 MeV and 0 - 200 keV for events coming within the first 100 ms after 
a $^{29}$Ne implant. $\gamma$-rays from $^{29}$Ne $\beta^-$, $\beta$-n decay 
and $^{29}$Na-$\beta^-$, $\beta$-n decay are indicated. The inset shows the 
$\beta$-$\gamma$-$\gamma$ coincidences gated by 72 keV transition.}
\label{fig1}
\end{center}
\end{figure}

The parent nuclei, $^{28,29}$Ne, were produced by fragmentation of a 140 
MeV/nucleon $^{48}$Ca$^{20+}$ beam in a 733 mg/cm$^2$  Be target located at the
 object position of the fragment separator, A1900, at the National 
Superconducting Cyclotron Laboratory (NSCL) at Michigan State University. A 
300 mg/cm$^2$ wedge-shaped Al degrader was used at the intermediate image of 
the A1900 to disperse the fragments according to their M/Z ratios. The 
transport of $^{28,29}$Ne ions was optimized in two settings of the A1900 
magnetic fields: 4.52 Tm and 4.39 Tm with momentum acceptance of 1\% for 
$^{28}$Ne and 4.69 Tm and 4.56 Tm with momentum acceptance of 2\% for 
$^{29}$Ne. A ``cocktail'' secondary beam was obtained in both cases. The yield 
of $^{29}$Ne was $\approx$0.14 particles/s/pnA. The fully stripped fragments 
were implanted in a double sided Si micro-strip detector (DSSD), part of the 
Beta Counting System (BCS) \cite {prisci} at NSCL. A 10 mm thick Al degrader 
was placed before the DSSD to ensure complete implantation within the DSSD. 
Fragments were unambiguously identified by a combination of multiple energy 
loss signals and time of flight. Each recorded event was tagged with an 
absolute time stamp generated by a free-running clock (30.5 $\mu$s repetition).
 Fragment-$\beta$ correlations were established in software, a low-energy 
$\beta$ event was correlated  to a high-energy implant event in a same or 
adjacent pixels of the DSSD. Light particles were vetoed by a scintillator at 
the end of BCS, increasing the fragment-$\beta$ correlation efficiency. 
The differences between the absolute time stamps of the correlated $\beta$ and 
implant events were histogrammed to generate a decay curve. To suppress 
background, implants were rejected if they were not followed by a $\beta$ event
 within a specified time period in the same (or neighboring) pixel or if a 
second implantation occured before a $\beta$ decay. A 100 ms time period was 
chosen for $^{28,29}$Ne decay analysis. The $\beta$-delayed $\gamma$ rays were 
detected using 12 detectors of the SEgmented Germanium Array (SeGA) 
\cite{mueller} arranged around the BCS. The Ge detectors were energy and 
efficiency calibrated using standard calibrated sources. 

The large $Q_\beta$ window ($>$10 MeV) for neutron rich nuclei and small 
neutron separation energies ($\sim$4 MeV) in daughter nuclei allow for the 
population of many excited bound and neutron-unbound states after $\beta^-$
decay.  The $\beta$-delayed one and two neutron emission probabilities, P$_n$ 
and P$_{2n}$, for $^{28,29}$Ne, have been extracted from the $\gamma$ 
activities of the grand-daughter nuclei. The values obtained for $^{29}$Ne are 
29 $\pm$ 7 \% and 4 $\pm$ 1 \% respectively. P$_n$ agrees well with the 
previous value of 27 $\pm$ 9 \% obtained by $\beta$-neutron coincidences \cite 
{reed}, whereas no previous measurement of P$_{2n}$ is reported. For $^{28}$Ne 
$\beta$-decay,  P$_n$ and P$_{2n}$ are 12 $\pm$ 1 \% and 3 $\pm$ 1\% 
respectively.  The half lives for the $\beta$-decay of $^{28,29}$Ne, extracted 
from a fitting of the decay curves, incorporating these neutron decay branches 
are 18.4 $\pm$ 0.5 ms and  13.8 $\pm$ 0.5 ms respectively, in agreement with 
reported values \cite {notoni}.

The energy spectra of $\beta$-delayed $\gamma$ rays emitted within 100 ms 
($\sim$5 half lives) of the arrival of an implant were generated for $^{28}$Ne 
and $^{29}$Ne. Parts of the spectrum for $^{29}$Ne decay are shown in Fig. 1, 
where transitions associated with the $\beta^-$ decay of $^{29}$Ne are 
identified.
\begin{figure}
\begin{center}
\includegraphics[scale=1.4]{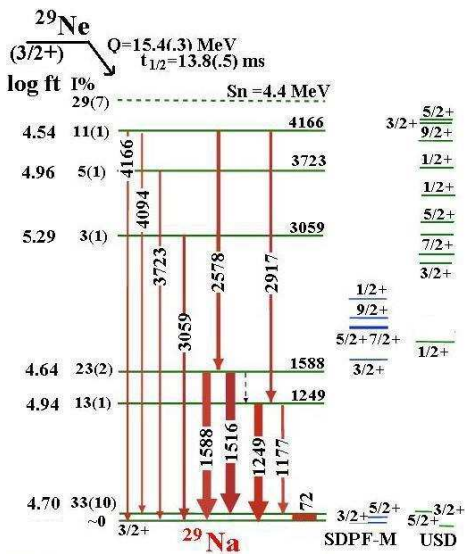}
\caption{(Color online) Proposed level scheme for $^{29}$Na populated following
 the $\beta$-decay of $^{29}$Ne. The absolute $\beta$-decay branching to each 
level per 100 decay is indicated along with the calculated log {\it f}t values.
 Shown on the right are shell model calculation with the USD and SDPF-M 
interactions.}
\label{fig2}
\end{center}
\end{figure}
Decay curves generated in coincidence with these $\gamma$ lines yielded 
consistent half lives, justifying their placement in the level 
scheme of $^{29}$Na, shown in Fig 2. The observation of pairs of lines, 1177 
keV (5\% $\pm$ 1\%) - 1249 keV (12\% $\pm$ 1\%)  and 1516 keV (16\% $\pm$ 2\%) 
- 1588 keV (11\% $\pm$ 2\%) differing by 72 keV and the observation of the 72 
keV (54\% $\pm$ 9\%) transition itself, confirms the first three excited 
states. The placement of the 1588 keV level is supported by the observed 
coincidence between the 72 keV and 1516 keV transitions (inset of Fig. 1).
  The other strong $\gamma$-rays, 2578 keV (5\% $\pm$ 1\%) and 2917 keV (3.5\% 
$\pm$ 0.5\%) depopulate the 4166 keV level. For $^{28}$Ne decay, 
$\beta\gamma\gamma$ coincidences were the main guidelines in generating the 
level scheme.  Coincidences for the 55 keV with 1076 keV, 2063 keV and 2659 keV
 transitions can be seen in Fig. 3, suggesting 4 levels at excitation energies 
of 55 keV, 1131 keV,  2118 keV and 2714 keV. The position of the 2714 keV state
 is corroborated by the 1583 keV - 1076 keV coincidences, while the 863 keV - 
1255 keV cascade generates a level at 1255 keV. The $\gamma$-$\gamma$ 
coincidences and energy and intensity sums allowed us  to establish the 
detailed level scheme for $^{28}$Na, shown in Fig. 4.  The absolute intensities
 for $\beta$ decay were calculated using the measured SeGA efficiency and the 
total number of $^{28,29}$Ne decay events obtained from the decay curves. The 
log {\it f}t values were calculated from these  $\beta$ decay intensities 
(ignoring the weak unobserved transitions) according to Ref. \cite 
{nndc}. 
 
\begin{figure}
\begin{center}
\includegraphics[scale=.35]{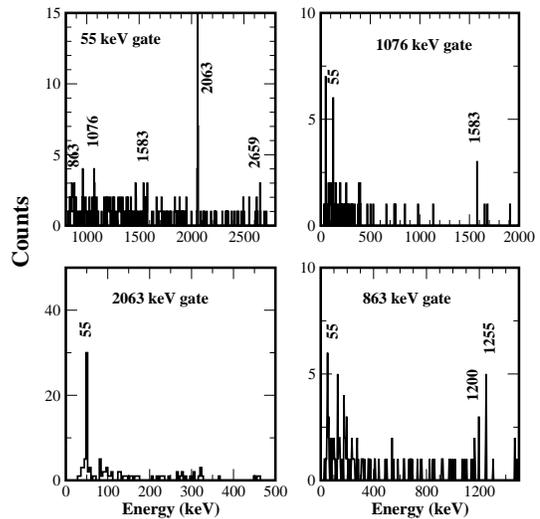}
\caption {The fragment-$\beta-\gamma-\gamma$ coincidences for 
$^{28}$Ne decay.}
\label{fig3}
\end{center}
\end{figure}

A comparison of the level schemes for $^{28,29}$Na from the current work with 
shell model calculations brings out interesting differences.  Clearly  the 
level structure for $^{29}$Na (Fig. 2) is in stark disagreement with the USD 
predictions  \cite {alex}. Apart from the ground state doublet (the order of 
which is reversed), the correspondence with the shell model predictions is not 
obvious. The experimental levels at 1249 keV and 1588 keV have large $\beta^-$
decay branches, implying spin assignments of 1/2$^+$, 3/2$^+$ or 5/2$^+$ 
($J^\pi$ of $^{29}$Ne ground state is calculated to be 3/2$^+$); however the 
USD calculations predict only one state (1/2$^+$) in this spin range below 2.8 
MeV which has a small predicted $\beta^-$ decay branch \cite {volya}. Also, 
within the USD shell model the almost degenerate states of the ground state 
doublet are expected to be equally populated via Gamow-Teller $\beta^-$ 
decay transition, this contradicts observation.  Allowing 
for excitations beyond the {\it sd} shell,  the MCSM calculation with SDPF-M 
interaction \cite {otsukanew}, predicts the correct ground state spin of 
$^{29}$Na (3/2$^+$ \cite {huber}). Four more states within the spin range, 
1/2$^+$- 5/2$^+$ are predicted below 2.5 MeV all with large probability of 
$2p2h$ excitations. The large $\beta$-decay branch to the 72 keV level makes 
it a likely candidate for the 5/2$^+_1$ state predicted as a member of the 
ground state doublet. The 3/2$^+_2$ (65\% of $2p2h$ contribution), 5/2$^+_2$ 
(78\% of $2p2h$ contribution) are good candidates for the 1249 keV and 1588 keV
 experimental levels. The better agreement seen between the experimental 
results and the MCSM calculations suggests that $fp$ intruder configurations 
play an important role in the low-energy level structure of $N=18$ $^{29}$Na. 
The present results also support a 3/2$^+$ assignment to the ground state of 
$^{29}$Ne, with strong $2p2h$ intruder mixing.  

For $^{28}$Na, the MCSM calculations predict low energy excited states which 
have almost pure $0p0h$ configurations in reasonable agreement with the USD 
calculations. Here intruder ($2p2h$) states involve the neutron excitation from
 the 1$s_{1/2}$ orbital, at a higher energy cost. The correspondence between 
the experimental level structure of $^{28}$Na and the calculations is quite 
good as demonstrated in Fig. 4. The ground state of $^{28}$Na was previously 
assigned 1$^+$ \cite {huber}, and the observation of a large $\beta^-$ decay 
branch confirms the 1$^+$ assignment.  The USD as well as MCSM calculations 
predict a $2^+$ ground state and a $1^+$ state with excitation energy less than
 100 keV. It appears this doublet is reversed in nature and the observed 55 keV
 first excited state is almost certainly the predicted $2^+$ level. Strong 
$\beta^-$ decay branches from $^{28}$Ne ($0^+$ ground state) have been observed
 to levels at  2118 keV and 2714 keV in the present work. These are likely the 
1$^+$ states in good agreement with the MCSM calculations, though the USD shell
 model predicts their energies a bit lower. The shell model (USD) predicts 
$\approx$75 \% of the total Gammow-Teller strength to go to the ground state 
and the first excited $1^+$ state \cite {volya} in close agreement with the 
experimentally calculated log {\it ft} values.  The 1255 keV state with 
negligible direct feeding most likely corresponds to the 1240(11) keV state 
seen by  B.V. Pritychenko {\it et al}, who assigned a spin of $J=2$ \cite 
{prity3}. Thus, there are candidates with closeby energies and log {\it ft} 
values in the USD shell model calculations, for the experimentally observed 
states, suggesting that $^{28}$Na can be described rather well with pure {\it 
sd} shell configurations without invoking mixing of intruder configurations.

\begin{figure}
\begin{center}
\includegraphics[scale=1.4]{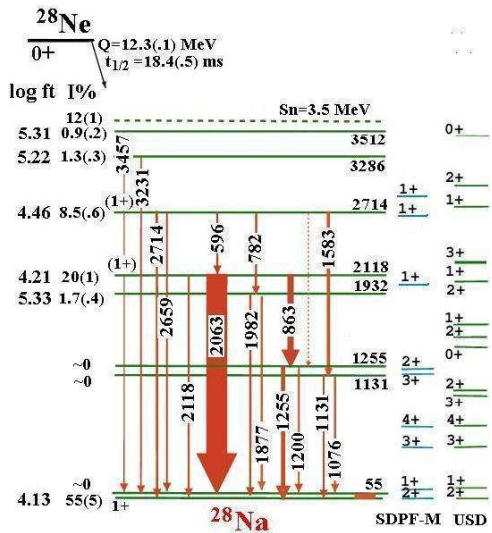}
\caption{(Color online) Proposed level scheme for $^{28}$Na populated following
 the $\beta^-$ decay of $^{28}$Ne. }
\label{fig4}
\end{center}
\end{figure}

The presence of states at low excitation energy which have dominant $2p2h$ 
configuration in $^{29}$Na points to the small $N=20$ shell gap allowing for 
such intruder states. As suggested by Otsuka {\it et al.,} \cite {otsuka}, the 
$N=20$ shell gap evolves as a function of Z, being the smallest for O. Study of
 the level schemes for $^{25-28}$Ne \cite {prity, azaiez} suggested that in the
 Ne isotopes, interference of intruders  already starts at $N=17$ and the 
neutron {\it fp} shell has to be included in the valence space in order to 
reproduce the data. However, the $N=18$, $^{30}$Mg \cite {30mg} shows better 
agreement with the shell model predictions including only the {\it sd} shell.  
For the case of neutron rich Na isotopes, we find that $^{28}$Na behaves nearly
 ``normally'', as discussed above,  but the $N=18$ isotope, $^{29}$Na, shows 
clear signatures of the influence of intruder configurations in its low energy 
level structure. Thus it is obvious that the  intruder configuration mixing 
becomes important  at $N=17,18,19$ as we traverse from Ne to Mg via Na. This is
 a direct experimental signature of the evolution of the underlying $N=20$ 
shell gap as a function of $Z$. 
 
In summary, a study of the $\beta^-$ decay of neutron-rich $^{28,29}$Ne is 
reported. The low-energy level structure for the  $N=17,18$ isotopes of Na have
 been established for the first time.  Comparison of the level schemes and 
$\beta^-$ branching ratios with theoretical shell model predictions, shows the 
influence of intruder configurations in the low-energy level structure of 
$^{29}$Na, but not in $^{28}$Na. For the Na isotopic chain, this is the first 
attempt to delineate the transition from the normal to intruder domination as 
seen in the low-energy level structure. The presence of states at low energy in
 $^{29}$Na with dominant intruder configuration is a signature of the smaller 
shell gap compared to that of stable nuclei. 

This work was supported by the NSF grants PHY-01-39950 and PHY-01-10253 and 
DoE grant DE-FG02-92ER40750. The authors thank the NSCL 
operations staff for the smooth conduct of the experiment.

\end{document}